\documentclass[12pt]{article}
\usepackage{amsmath,amssymb}
\usepackage{graphicx}
\newcommand{\eqdef}{\stackrel{\text{def}}{=}}
\newcommand{\n}{\nonumber\\}
\newcommand{\bm}{\boldsymbol}
\newcommand{\ignore}[1]{}
\numberwithin{equation}{section}
\newcommand{\Romannumeral}[1]{\uppercase\expandafter{\romannumeral#1}}

\newcommand{\II}{\text{\Romannumeral{2}}}
\newtheorem{theo}{\bf Theorem}
\newtheorem{conj}[theo]{\bf Conjecture}

\newtheorem{rema}{\bf Remark}

\allowdisplaybreaks[4]

\begin{document}

\baselineskip=20pt

\newcommand{\preprint}{
    \begin{flushright}\normalsize \sf
     DPSU-14-2\\
   \end{flushright}}
\newcommand{\Title}[1]{{\baselineskip=26pt
   \begin{center} \Large \bf #1 \\ \ \\ \end{center}}}
\newcommand{\Author}{\begin{center}
   \large \bf Satoru Odake${}^a$ and Ryu Sasaki${}^{a,b}$ \end{center}}
\newcommand{\Address}{\begin{center}
     $^a$ Department of Physics, Shinshu University,\\
     Matsumoto 390-8621, Japan\\
     ${}^b$ Department of Physics,\\
     National Taiwan University, Taipei 10617, Taiwan
    \end{center}}
\newcommand{\Accepted}[1]{\begin{center}
   {\large \sf #1}\\ \vspace{1mm}{\small \sf Accepted for Publication}
   \end{center}}

\preprint
\thispagestyle{empty}

\Title{Reflectionless Potentials for Difference Schr\"odinger Equations}

\Author

\Address
\vspace{1cm}

\begin{abstract}
As a part of the program `discrete quantum mechanics,' we present general
reflectionless potentials for difference Schr\"odinger equations with pure
imaginary shifts. By combining contiguous integer wave number reflectionless
potentials, we construct the discrete analogues of the $h(h+1)/\cosh^2x$
potential with the integer $h$, which belong to the recently constructed
families of solvable dynamics having the $q$-ultraspherical polynomials with
$|q|=1$ as the main part of the eigenfunctions.
For the general ($h\in\mathbb{R}_{>0}$) scattering theory for these potentials,
we need the connection formulas for the basic hypergeometric function
${}_2\phi_1(\genfrac{}{}{0pt}{}{a,\,b}{c}|q\,;z)$ with $|q|=1$,
which is not known. The connection formulas are expected to contain the quantum
dilogarithm functions as the $|q|=1$ counterparts of the $q$-gamma functions.
We propose a conjecture of the connection formula of the ${}_2\phi_1$ function
with $|q|=1$. Based on the conjecture, we derive the transmission and
reflection amplitudes, which have all the desirable properties. They provide
a strong support to the conjectured connection formula.
\end{abstract}


\section{Introduction}
\label{sec:intro}

It is well known that the general {\em reflectionless potentials\/} \cite{KM}
in ordinary Quantum Mechanics (QM for short) can be constructed from the
vacuum by repeated application of Darboux transformations in terms of the
exponential seed solutions $e^{k_jx}+\tilde{c}_je^{-k_jx}$ ($k_j>0$,
$\tilde{c}_j\in\mathbb{R}$) \cite{matv-sal,refless}.
This fact also explains that the reflectionless potentials are
{\em exactly solvable\/}, that is, all the eigenvalues $\{-k_j^2\}$ and
the corresponding eigenfunctions are obtained explicitly.
In this case, all the bound state eigenfunctions are elementary functions
expressed as a ratio of determinants.
It is also known that the typical {\em exactly solvable potential\/}
$h(h+1)/\cosh^2x$ \cite{infhull,susyqm}, with {\em integer\/}
$h\in\mathbb{Z}_{>0}$ (reflectionless case), can be constructed by special
combinations of the integer wave number reflectionless potentials,
$k_j=j$, $\tilde{c}_j=(-1)^{j-1}$ $(j=1,\ldots,h)$.
For this potential with generic $h$, besides the eigenpolynomial, which is
the Jacobi polynomial, the scattering amplitudes are exactly calculable
through the {\em connection formulas of the Gaussian hypergeometric
functions\/} \cite{KS,HLS}.
 
The motivation of this paper is to present the parallel results in
{\em discrete Quantum Mechanics with pure imaginary shifts\/} \cite{os13,os24}.
That is,
(\romannumeral1)
To construct general reflectionless potentials by multiple Darboux
transformations \cite{os15,gos,os27} in terms of the exponential seed
solutions of the vacuum. The corresponding discrete eigenvalues and
eigenfunctions are obtained simultaneously. That is, the reflectionless
potentials are {\em exactly solvable\/}.
(\romannumeral2)
By combining special exponential seed solutions corresponding to contiguous
integer wave numbers, the discrete counterpart of the $h(h+1)/\cosh^2x$
potential is constructed. The obtained potentials are identified as one of
the exactly solvable systems introduced in our recent work \cite{os32} having
the sinusoidal coordinates $\eta(x)=\sinh x$ and $|q|=1$.
For the generic couplings, the ground state wave functions consist of
{\em quantum dilogarithm\/} functions \cite{qdilog}. The excited states wave
functions are described by the {\em $q$-ultraspherical polynomials\/} and/or
Heine's hypergeometric functions ${}_2\phi_1$ \cite{gasper}.
For the special couplings realizing the reflectionless potentials, the
quantum dilogarithms are shown to degenerate into elementary functions.
(\romannumeral3)
We propose {\em a conjecture of the connection formula for the basic
hypergeometric function ${}_2\phi_1$ with $|q|=1$}, built upon the empirical
correspondence between the infinite $q$-shifted factorials and the quantum
dilogarithms. The transmission and reflection amplitudes for generic couplings
are derived based on the conjectured connection formula for the basic
hypergeometric function ${}_2\phi_1(\genfrac{}{}{0pt}{}{a,\,b}{c}|q\,;z)$
with $|q|=1$. The obtained scattering amplitudes have all the desirable
properties and they provide a strong support for the conjectured connection
formula.

Reflectionless potentials in discrete QM with real shifts were reported in
\cite{SZ}.

\medskip

This paper is organized as follows.
In section two the general reflectionless potentials in discrete QM with
pure imaginary shifts are derived from the trivial potential ($V(x)\equiv1$)
by multiple Darboux transformations in terms of exponential seed solutions.
The explicit forms of the discrete eigenfunctions and the transmission
amplitude are obtained.
The discrete counterpart of the $1/\cosh^2x$ potential is constructed in
section three. A brief summary of the construction of the $1/\cosh^2x$
potential by the special combination of reflectionless potentials in
ordinary QM is given in \S\,\ref{sec:ordsoliton}.
The derivation of the discrete counterpart is done in two steps.
The reflectionless case is derived directly from the special combinations
of discrete reflectionless potentials in \S\,\ref{sec:disreflesssoliton}.
It is identified with a special case of the discrete exactly solvable
systems with the sinusoidal coordinate $\eta(x)=\sinh x$ introduced earlier
\cite{os32}. The properties of the generic case are explored in some detail
in \S\,\ref{sec:disgensoliton}. A conjectured connection formula for the
basic hypergeometric function with the base $|q|=1$ is presented in
\S\,\ref{sec:conject}.
The scattering amplitudes are derived and their properties are examined.
The final section is for a summary and comments.
A few supporting evidences for the conjectured connection formula
of the basic hypergeometric functions and for the replacement rule between
the infinite $q$-shifted factorial and the quantum dilogarithm function
are given in Appendix.

\section{Reflectionless Potentials}
\label{sec:refless}

Here we derive the general reflectionless potentials in discrete QM with
pure imaginary shifts.
Let us start with the basic formulation.
In discrete QM the momentum operator $p=-i\frac{d}{dx}$ enters in the
Hamiltonian in the exponentiated forms $e^{\pm \gamma p}$
($\gamma\in\mathbb{R}$), which give pure imaginary shifts to the wave function
\begin{equation*}
  e^{\pm\gamma p}\psi(x)=\psi(x\mp i\gamma).
\end{equation*}
The Hamiltonian $\mathcal{H}$ depends on an analytic function of $x$, $V(x)$
and it has a factorised form:
\begin{align}
  \mathcal{H}
  &=\sqrt{V(x)V^*(x-i\gamma)}\,e^{\gamma p}
  +\sqrt{V^*(x)V(x+i\gamma)}\,e^{-\gamma p}-V(x)-V^*(x)
  \label{H_idQM}\\
  &=\mathcal{A}^{\dagger}\mathcal{A},\\
  \mathcal{A}&=i\bigl(e^{\frac{\gamma}{2}p}\sqrt{V^*(x)}
  -e^{-\frac{\gamma}{2}p}\sqrt{V(x)}\,\bigr),\quad
  \mathcal{A}^{\dagger}=-i\bigl(\sqrt{V(x)}\,e^{\frac{\gamma}{2}p}
  -\sqrt{V^*(x)}\,e^{-\frac{\gamma}{2}p}\bigr),
\end{align}
where $\gamma\in\mathbb{R}_{\neq 0}$ is a real parameter and we assume
$\gamma>0$ in the following.
The $*$-operation on an analytic function $f(x)=\sum_na_nx^n$
($a_n\in\mathbb{C}$) is defined by $f^*(x)\eqdef f(x^*)^*=\sum_na_n^*x^n$,
in which $a_n^*$ is the complex conjugation of $a_n$.
The trivial choice $V(x)\equiv1$ gives the free Hamiltonian $\mathcal{H}_0$
together with the exponential and wave solutions:
\begin{align}
  &\mathcal{H}_0=e^{\gamma p}+e^{-\gamma p}-2,\quad\,p=-i\frac{d}{dx},
  \label{H0_idQM}\\
  &\mathcal{H}_0e^{\pm kx}=\tilde{\mathcal{E}}_ke^{\pm kx},\qquad
  \tilde{\mathcal{E}}_k\eqdef-4\sin^2\frac{k\gamma}2<0,\\
  &\mathcal{H}_0e^{\pm ikx}=\mathcal{E}_k^{\text{s}}e^{\pm ikx},\ \,\quad
  \mathcal{E}_k^{\text{s}}\eqdef 4\sinh^2\frac{k\gamma}2>0.
\end{align}
For $\tilde{\mathcal{E}}_k$, we restrict $k$ in the range
$0<k\leq\frac{\pi}{\gamma}$.
Let us call $e^{ikx}$ the right moving wave and $e^{-ikx}$ the left moving wave.
In the $\gamma\to 0$ limit, these quantities reduce to the well-known
counterparts in ordinary QM:
\begin{equation}
  \lim_{\gamma\to 0}\gamma^{-2}\mathcal{H}_0=p^2,\quad
  \lim_{\gamma\to 0}\gamma^{-2}\tilde{\mathcal{E}}_k=-k^2,\quad
  \lim_{\gamma\to 0}\gamma^{-2}\mathcal{E}^{\text{s}}_k=k^2,
\end{equation}
and the range of $k$ becomes simply $0<k$.

We deform $\mathcal{H}_0$ by the $N$-step Darboux transformations
\cite{os15,gos,os27} in terms of  the following {\em exponential seed
solutions\/} $\psi_1(x)$, $\psi_2(x),\ldots,\psi_N(x)$ in this order,
\begin{align}
  &\psi_j(x)=e^{k_jx}+\tilde{c}_je^{-k_jx}=\psi_j^*(x),\quad
  0<k_1<k_2<\cdots<k_N\leq\tfrac{\pi}{\gamma},\ \ (-1)^{j-1}\tilde{c}_j>0,
  \label{psij_idQM}\\
  &\mathcal{H}_0\psi_j(x)=\tilde{\mathcal{E}}_{k_j}\psi_j(x).
\end{align}
Since the inverse of these seed solutions are locally square integrable
at $x=\pm\infty$, these transformations are state adding transformations
and $N$ eigenstates are added if the resulted Hamiltonian is non-singular.
First we rewrite $\mathcal{H}_0$ as
\begin{equation}
  \mathcal{H}_0=\hat{\mathcal{A}}_1^{\dagger}\hat{\mathcal{A}}_1
  +\tilde{\mathcal{E}}_{k_1},\quad
  \hat{\mathcal{A}}_1\eqdef i\bigl(e^{\frac{\gamma}{2}p}\sqrt{\hat{V}_1^*(x)}
  -e^{-\frac{\gamma}{2}p}\sqrt{\hat{V}_1(x)}\,\bigr),
  \ \ \hat{V}_1(x)\eqdef\frac{\psi_1(x-i\gamma)}{\psi_1(x)}.
\end{equation}
Then we repeat the Darboux transformations:
\begin{align}
  &\mathcal{H}_{12\ldots s}\eqdef
  \hat{\mathcal{A}}_{12\ldots s}\hat{\mathcal{A}}_{12\ldots s}^{\dagger}
  +\tilde{\mathcal{E}}_{k_s}
  =\hat{\mathcal{A}}_{12\ldots s\,s+1}^{\dagger}
  \hat{\mathcal{A}}_{12\ldots s\,s+1}
  +\tilde{\mathcal{E}}_{k_{s+1}},\\
  &\hat{\mathcal{A}}_{12\ldots s}\eqdef
  i\bigl(e^{\frac{\gamma}{2}p}\sqrt{\hat{V}_{12\ldots s}^*(x)}
  -e^{-\frac{\gamma}{2}p}\sqrt{\hat{V}_{12\ldots s}(x)}\,\bigr),\\
  &\hat{V}_{12\ldots s}(x)\eqdef
  \frac{\text{W}_{\gamma}[\psi_1,\psi_2,\ldots,\psi_{s-1}](x+i\frac{\gamma}{2})}
  {\text{W}_{\gamma}[\psi_1,\psi_2,\ldots,\psi_{s-1}](x-i\frac{\gamma}{2})}
  \frac{\text{W}_{\gamma}[\psi_1,\psi_2,\ldots,\psi_s](x-i\gamma)}
  {\text{W}_{\gamma}[\psi_1,\psi_2,\ldots,\psi_s](x)}.
\end{align}
The Casoratian of functions $\{f_j(x)\}_{j=1}^n$ is defined by
\begin{equation}
  \text{W}_{\gamma}[f_1,\ldots,f_n](x)
  \eqdef i^{\frac12n(n-1)}
  \det\Bigl(f_k\bigl(x^{(n)}_j\bigr)\Bigr)_{1\leq j,k\leq n},\quad
  x_j^{(n)}\eqdef x+i(\tfrac{n+1}{2}-j)\gamma,
  \label{Wgammadef}
\end{equation}
($\text{W}_{\gamma}[\cdot](x)=1$ for $n=0$).
The final step Hamiltonian
$\mathcal{H}^{[N]}\eqdef\mathcal{H}_{12\ldots N}$,
\begin{equation}
  \mathcal{H}^{[N]}=\hat{\mathcal{A}}^{[N]}\hat{\mathcal{A}}^{[N]\,\dagger}
  +\tilde{\mathcal{E}}_{k_N},\quad
  \hat{\mathcal{A}}^{[N]}\eqdef\hat{\mathcal{A}}_{12\ldots N},\quad
  \hat{V}^{[N]}(x)\eqdef\hat{V}_{12\ldots N}(x),
\end{equation}
is rewritten as, by setting
$\mathcal{A}^{[N]}\eqdef-\hat{\mathcal{A}}^{[N]\,\dagger}$,
\begin{align}
  &\mathcal{H}^{[N]}=\mathcal{A}^{[N]\,\dagger}\mathcal{A}^{[N]}
  +\tilde{\mathcal{E}}_{k_N},\quad
  \mathcal{A}^{[N]}\eqdef
  i\bigl(e^{\frac{\gamma}{2}p}\sqrt{V^{[N]\,*}(x)}
  -e^{-\frac{\gamma}{2}p}\sqrt{V^{[N]}(x)}\,\bigr),\\
  &V^{[N]}(x)\eqdef\hat{V}^{[N]\,*}(x-i\tfrac{\gamma}{2})=
  \frac{\text{W}_{\gamma}[\psi_1,\ldots,\psi_{N-1}](x-i\gamma)}
  {\text{W}_{\gamma}[\psi_1,\ldots,\psi_{N-1}](x)}
  \frac{\text{W}_{\gamma}[\psi_1,\ldots,\psi_N](x+i\frac{\gamma}{2})}
  {\text{W}_{\gamma}[\psi_1,\ldots,\psi_N](x-i\frac{\gamma}{2})}.
  \label{VN}
\end{align}

The Casoratian of exponential functions is
\begin{equation}
  \text{W}_{\gamma}[e^{k_1x},e^{k_2x},\ldots,e^{k_nx}](x)
  =\prod_{1\leq i<j\leq n}\!\!2\sin\tfrac{\gamma}{2}(k_j-k_i)
  \cdot e^{\sum_{j=1}^nk_jx}.
  \label{Wexp_idQM}
\end{equation}
By the choice of the signs of the constants $(-1)^{j-1}\tilde{c}_j>0$
($j=1,\ldots,N$), it is easy to see that the Casoratians are positive for
real $x$:
\begin{equation}
  \text{W}_{\gamma}[\psi_1,\ldots,\psi_j](x)>0,\quad
  -\infty<x<\infty,\quad j=1,\ldots,N.
\end{equation}
This means that Hamiltonian $\mathcal{H}^{[N]}$ is non-singular.\footnote{
For the hermiticity of the Hamiltonian, however, we have to
show that $u_N(x)$ \eqref{uN_idQM} has no zeros in the strip
$|\text{Im}\,x|<\frac12\gamma$ in the complex plane (for example, see
\cite{os32}).
}
Its eigenstates $\Phi^{[N]}_j(x)$ ($j=1,2,\ldots,N$) and the right moving wave
$\Psi^{[N]}_k(x)$ ($k>0$) are (for example, see \cite{os30}):
\begin{align}
  &\mathcal{H}^{[N]}\Phi^{[N]}_j(x)
  =\tilde{\mathcal{E}}_{k_j}\Phi^{[N]}_j(x),\quad
  \Phi^{[N]}_j(x)
  =\frac{\text{W}_{\gamma}[\psi_1,\ldots,\breve{\psi}_j,\ldots,\psi_N](x)}
  {\bigl(\text{W}_{\gamma}[\psi_1,\ldots,\psi_N](x-i\frac{\gamma}{2})
  \text{W}_{\gamma}[\psi_1,\ldots,\psi_N](x+i\frac{\gamma}{2})
  \bigr)^{\frac12}},
  \label{Phij_idQM}\\
  &\mathcal{H}^{[N]}\Psi^{[N]}_k(x)
  =\mathcal{E}^{\text{s}}_k\Psi^{[N]}_k(x),\quad
  \Psi^{[N]}_k(x)=\frac{\text{W}_{\gamma}[\psi_1,\ldots,\psi_N,e^{ikx}](x)}
  {\bigl(\text{W}_{\gamma}[\psi_1,\ldots,\psi_N](x-i\frac{\gamma}{2})
  \text{W}_{\gamma}[\psi_1,\ldots,\psi_N](x+i\frac{\gamma}{2})
  \bigr)^{\frac12}}.
  \label{Psik_idQM}
\end{align}
In \eqref{Phij_idQM} $\breve{\psi_j}$ means that $\psi_j$ is excluded from the
Casoratian.
By using this, the asymptotic forms of $\Phi^{[N]}_j(x)$ and
$\Psi^{[N]}_k(x)$ at $x\to\infty$ are:
\begin{align}
  \Phi^{[N]}_j(x)&\approx
  \frac{\text{W}_{\gamma}[e^{k_1x},\ldots,\breve{e^{k_jx}},\ldots e^{k_Nx}](x)}
  {\bigl(\text{W}_{\gamma}[e^{k_1x},\ldots,e^{k_Nx}](x-i\frac{\gamma}{2})
  \text{W}_{\gamma}[e^{k_1x},\ldots,e^{k_Nx}](x+i\frac{\gamma}{2})
  \bigr)^{\frac12}}\n
  &=(-1)^{j-1}\prod_{\genfrac{}{}{0pt}{}{i=1}{i\neq j}}^N
  \bigl(2\sin\tfrac{\gamma}{2}(k_i-k_j)\bigr)^{-1}\cdot e^{-k_jx},\\
  \Psi^{[N]}_k(x)&\approx
  \frac{\text{W}_{\gamma}[e^{k_1x},\ldots,e^{k_Nx},e^{ikx}](x)}
  {\bigl(\text{W}_{\gamma}[e^{k_1x},\ldots,e^{k_Nx}](x-i\frac{\gamma}{2})
  \text{W}_{\gamma}[e^{k_1x},\ldots,e^{k_Nx}](x+i\frac{\gamma}{2})
  \bigr)^{\frac12}}\n
  &=\prod_{j=1}^N2\sin\tfrac{\gamma}{2}(ik-k_j)\cdot e^{ikx},
\end{align}
and the asymptotic forms at $x\to-\infty$ are:
\begin{align}
  \Phi^{[N]}_j(x)&\approx
  \frac{\text{W}_{\gamma}[\tilde{c}_1e^{-k_1x},\ldots,
  \tilde{c}_j\breve{e^{-k_jx}},\ldots\tilde{c}_Ne^{-k_Nx}](x)}
  {\bigl(\text{W}_{\gamma}[\tilde{c}_1e^{-k_1x},\ldots,\tilde{c}_Ne^{-k_Nx}]
  (x-i\frac{\gamma}{2})
  \text{W}_{\gamma}[\tilde{c}_1e^{-k_1x},\ldots,\tilde{c}_Ne^{-k_Nx}]
  (x+i\frac{\gamma}{2})\bigr)^{\frac12}}\n
  &=(-1)^{j-1}\tilde{c}_j^{-1}
  \prod_{\genfrac{}{}{0pt}{}{i=1}{i\neq j}}^N
  \bigl(2\sin\tfrac{\gamma}{2}(k_j-k_i)\bigr)^{-1}\cdot e^{k_jx},\\
  \Psi^{[N]}_k(x)&\approx
  \frac{\text{W}_{\gamma}[\tilde{c}_1e^{-k_1x},\ldots,
  \tilde{c}_Ne^{-k_Nx},e^{ikx}](x)}
  {\bigl(\text{W}_{\gamma}[\tilde{c}_1e^{-k_1x},\ldots,\tilde{c}_Ne^{-k_Nx}]
  (x-i\frac{\gamma}{2})
  \text{W}_{\gamma}[\tilde{c}_1e^{-k_1x},\ldots,\tilde{c}_Ne^{-k_Nx}]
  (x+i\frac{\gamma}{2})\bigr)^{\frac12}}\n
  &=\prod_{j=1}^N2\sin\tfrac{\gamma}{2}(ik+k_j)\cdot e^{ikx}.
\end{align}
{}From these, $\Phi^{[N]}_j(x)$ is indeed square integrable, and the
transmission and reflection amplitudes of $\Psi^{[N]}_k(x)$ are
\begin{equation}
  t^{[N]}(k)=\prod_{j=1}^N\frac{\sin\frac{\gamma}{2}(ik-k_j)}
  {\sin\frac{\gamma}{2}(ik+k_j)}
  =\prod_{j=1}^N\frac{\sinh\frac{\gamma}{2}(k+ik_j)}
  {\sinh\frac{\gamma}{2}(k-ik_j)},\quad r^{[N]}(k)=0.
  \label{tkrk_idQM}
\end{equation}
Thus we have established that  {\em the general reflectionless potentials
are exactly solvable including the scattering problem\/}.
We have
\begin{equation}
  \mathcal{H}^{[N]}\phi^{[N]}_n(x)=\mathcal{E}^{[N]}_n\phi^{[N]}_n(x)
  \ \ (n=0,1,\ldots,N-1),\quad
  \mathcal{E}^{[N]}_0<\mathcal{E}^{[N]}_1<\cdots<\mathcal{E}^{[N]}_{N-1}<0,
  \label{HNphiNn=}
\end{equation}
in which $\phi^{[N]}_n(x)$ and $\mathcal{E}^{[N]}_n$ are defined by
\begin{align}
  \phi^{[N]}_n(x)&\eqdef\text{const}\times\Phi^{[N]}_{N-n}(x),\quad
  \mathcal{E}^{[N]}_n\eqdef\tilde{\mathcal{E}}_{k_{N-n}}\quad
  (n=0,1,\ldots,N-1).
  \label{phiN=PhiN_2}
\end{align}
Note that the Hamiltonian $\mathcal{H}^{[N]}$ has $N$ eigenstates of
arbitrary eigenvalues $\mathcal{E}^{[N]}_n$.
The potential function $V^{[N]}(x)$ and the eigenfunction $\Phi^{[N]}_j(x)$
are rational functions of $e^{k_jx}$.

\bigskip
In the rest of this section, we show that some quantities related with the
discrete reflectionless potentials have similar expressions to the
counterparts in ordinary QM, {\em i.e.\/} the profile of the KdV solitons
\cite{KM,refless,hirota}.
By introducing $u_N(x)$ by
\begin{equation}
  \text{W}_{\gamma}[\psi_1,\ldots,\psi_N](x)
  =\!\!\prod_{1\leq i<j\leq N}\!\!2\sin\tfrac{\gamma}{2}(k_j-k_i)\cdot
  e^{\sum_{j=1}^Nk_jx}u_N(x),
  \label{uN_idQM}
\end{equation}
($\Rightarrow u_N(x)=u_N^*(x)$),
the potential function $V^{[N]}(x)$ \eqref{VN} is expressed as
\begin{equation}
  V^{[N]}(x)=e^{i\gamma k_N}\frac{u_{N-1}(x-i\gamma)}{u_{N-1}(x)}
  \frac{u_N(x+i\frac{\gamma}{2})}{u_N(x-i\frac{\gamma}{2})}.
  \label{VN2}
\end{equation}
This function $u_N(x)$ \eqref{uN_idQM} can be expressed in a determinant form,
\begin{equation}
  u_N(x)=\det A_N(x),\quad
  \bigl(A_N(x)\bigr)_{mn}\eqdef
  \delta_{mn}+\frac{c_me^{-(k_m+k_n)x}}{\sin\frac{\gamma}{2}(k_m+k_n)}
  \ \ (m,n=1,2,\ldots,N).
  \label{uNdet_idQM}
\end{equation}
Here $\tilde{c}_j$ and $c_j$ are related by
\begin{equation}
  \tilde{c}_j=\frac{c_j}{\sin\gamma k_j}
  \prod_{\genfrac{}{}{0pt}{}{i=1}{i\neq j}}^N
  \frac{\sin\frac{\gamma}{2}(k_i-k_j)}{\sin\frac{\gamma}{2}(k_i+k_j)},
  \label{ctjcj_idQM}
\end{equation}
and the condition $(-1)^{j-1}\tilde{c}_j>0$ means $c_j>0$.
This determinant has the following expansion,
\begin{equation}
  u_N(x)=\!\!\!\sum_{\mu_1,\ldots,\mu_N=0}^1\!\!\!\!
  \exp\Bigl(\sum_{j=1}^N\mu_j\eta_j
  +\!\!\!\sum_{1\leq i<j\leq N}\!\!a_{ij}\mu_i\mu_j\Bigr),
  \ \ e^{\eta_j}=\frac{c_je^{-2k_jx}}{\sin\gamma k_j},
  \ \ e^{a_{ij}}=\Bigl(\frac{\sin\frac{\gamma}{2}(k_i-k_j)}
  {\sin\frac{\gamma}{2}(k_i+k_j)}\Bigr)^2,
  \label{uNexp_idQM}
\end{equation}
which is manifestly positive for real $x$ and positive
$k_j(<\frac{\pi}{\gamma})$ and $c_j$.
We define $u_{N,j}(x)$ ($j=1,\ldots,N$) by
\begin{equation}
  \text{W}_{\gamma}[\psi_1,\ldots,\breve{\psi}_j,\ldots,\psi_N](x)
  \!=\!\!\!\prod_{\genfrac{}{}{0pt}{}{1\leq i<l\leq N}{i,l\neq j}}
  \!\!\!2\sin\tfrac{\gamma}{2}(k_l-k_i)\cdot
  e^{\sum_{i=1}^Nk_ix-k_jx}u_{N,j}(x),
  \label{uNj_idQM}
\end{equation}
($\Rightarrow u_{N,j}(x)=u_{N,j}^*(x)$).
Then the eigenfunctions $\Phi^{[N]}_j(x)$ \eqref{Phij_idQM} are expressed as
\begin{equation}
  \Phi^{[N]}_j(x)=(-1)^{j-1}
  \prod_{\genfrac{}{}{0pt}{}{i=1}{i\neq j}}^N
  \bigl(2\sin\tfrac{\gamma}{2}(k_i-k_j)\bigr)^{-1}\cdot e^{-k_jx}
  \frac{u_{N,j}(x)}{\sqrt{u_N(x-i\frac{\gamma}{2})u_N(x+i\frac{\gamma}{2})}}.
\end{equation}
This $u_{N,j}(x)$ is written as a determinant by the replacement
$c_m\to\frac{\sin\frac{\gamma}{2}(k_j-k_m)}{\sin\frac{\gamma}{2}(k_j+k_m)}c_m$
($m=1,\ldots,N$) in \eqref{uNdet_idQM},
\begin{align}
  &u_{N,j}(x)=\det A_{N,j}(x),\n
  &\bigl(A_{N,j}(x)\bigr)_{mn}\eqdef\delta_{mn}
  +\frac{\sin\frac{\gamma}{2}(k_j-k_m)}{\sin\frac{\gamma}{2}(k_j+k_m)}
  \frac{c_me^{-(k_m+k_n)x}}{\sin\frac{\gamma}{2}(k_m+k_n)}
  \ \ (m,n=1,2,\ldots,N).
  \label{uNjdet_idQM}
\end{align}

In the $\gamma\to 0$ limit, the Casoratian reduces to the Wronskian,
\begin{equation}
 \lim_{\gamma\to 0}\gamma^{-\frac12n(n-1)}
 \text{W}_{\gamma}[f_1,\ldots,f_n](x)=\text{W}[f_1,\ldots,f_n](x),
\end{equation}
where $f_j(x)$'s are assumed to be independent of $\gamma$.
In the $\gamma\to 0$ limit, under the assumption that $k_j$ and $\tilde{c}_j$
are independent of $\gamma$ (\eqref{ctjcj_idQM} implies $c_j=O(\gamma)$),
the various quantities (with appropriate overall rescalings) in this section
reduce to the counterparts in ordinary quantum mechanics.
For example, the reflectionless potential is
\begin{align}
  U^{[N]}(x)&=-2\partial_x^2\log \mathfrak{u}_N(x),
  \label{UN2}\\
  \mathfrak{u}_N(x)&=\det \mathfrak{A}_N(x),\quad
  \bigl(\mathfrak{A}_N(x)\bigr)_{mn}\eqdef
  \delta_{mn}+\frac{\mathfrak{c}_me^{-(k_m+k_n)x}}{k_m+k_n}
  \ \ (m,n=1,2,\ldots,N).
  \label{uNdet_oQM}
\end{align}
Although the Hamiltonians are related directly, the relation between the
potential function $V^{[N]}(x)$ and the potential $U^{[N]}(x)$ is indirect.
The quantity related to $U^{[N]}(x)$ more directly is the potential function
$\mathcal{U}_N(x)$,
\begin{align}
  \mathcal{U}_N(x)&\eqdef
  \sqrt{V^{[N]}(x+i\tfrac{\gamma}{2})V^{[N]\,*}(x-i\tfrac{\gamma}{2})}
  =\mathcal{U}^{*}_N(x)
  \label{cUN}\\
  &=\frac{\sqrt{\text{W}_{\gamma}[\psi_1,\ldots,\psi_N](x+i\gamma)
  \text{W}_{\gamma}[\psi_1,\ldots,\psi_N](x-i\gamma)}}
  {\text{W}_{\gamma}[\psi_1,\ldots,\psi_N](x)}
  =\frac{\sqrt{u_N(x+i\gamma)u_N(x-i\gamma)}}{u_N(x)},
  \nonumber
\end{align}
whose limit is
\begin{equation}
  \lim_{\gamma\to 0}\bigl(\tfrac{2}{\gamma}\bigr)^2
  \bigl(\mathcal{U}_N(x)-1\bigr)
  =U^{[N]}(x).
\end{equation}

\bigskip
It is well known that the reflectionless potential in ordinary QM \eqref{UN2}
with a proper time dependence, that is,
\begin{equation}
  \mathfrak{c}_m\to\mathfrak{c}_me^{8k_m^3t}\quad(m=1,\ldots,N),
  \label{cjtime}
\end{equation}
gives the $N$-soliton solution of the KdV equation
\cite{hirota,matv-sal,refless}.
It is an interesting challenge to introduce an appropriate time dependence
to $\mathcal{U}_N(x)$, \eqref{cUN} so that $\mathcal{U}_N(x;t)$ satisfies
certain deformation of the KdV equation.

\section{Discrete Counterpart of $\bm{1/\cosh^2x}$ Potential}
\label{sec:soliton}

\subsection{$\bm{1/\cosh^2x}$ potential in ordinary QM}
\label{sec:ordsoliton}

First let us summarise fundamental properties of the $1/\cosh^2x$ potential
in ordinary QM. It is an exactly solvable system with finitely many discrete
eigenlevels and its scattering problem is also exactly solvable.

It is well known that the reflectionless $1/\cosh^2x$ potential is obtained
from the trivial potential $U(x)\equiv0$ by multiple Darboux transformations
in terms of a very special choice of the exponential seed solutions
\eqref{psij_idQM}
\begin{equation}
  k_j=j,\quad\tilde{c}_j=(-1)^{j-1}.
  \label{solitonkjctj}
\end{equation}
In other words, the potential \eqref{UN2} with
\begin{equation}
  k_j=j,\quad
  {\mathfrak{c}}_j=\tilde{c}_j2k_j
  \prod_{\genfrac{}{}{0pt}{}{i=1}{i\neq j}}^N\frac{k_i+k_j}{k_i-k_j}
  =\frac{(N+j)!}{j!(j-1)!(N-j)!}\quad(j=1,\ldots,N),
  \label{ordsolitonkjctj}
\end{equation}
gives rise to
\begin{equation}
  \mathfrak{U}^{[N]}(x)=-\frac{N(N+1)}{\cosh^2x},\quad
  \mathfrak{u}_N(x)=e^{-N(N+1)x}(1+e^{2x})^{\frac12N(N+1)},
  \label{solitonUN}
\end{equation}
together with its $n$-th discrete eigenfunction
\begin{equation}
  \phi_n(x)=(\cosh x)^{n-N}P_n^{(N-n,N-n)}(\tanh x),
\end{equation}
where $P^{(\alpha,\beta)}_n(\eta)$ is the Jacobi polynomial.
To be more precise, the above choice of $k_j$ and $\tilde{c}_j$
\eqref{solitonkjctj} gives rise to the Wronskians of the seed functions
$\psi_1,\ldots,\psi_N$ \eqref{psij_idQM}:
\begin{align}
  &\text{W}[\psi_1,\ldots,\psi_N](x)
  =\prod_{j=1}^{N-1}j^{N-j}\cdot(2\cosh x)^{\frac12N(N+1)},\\
  &\frac{\text{W}[\psi_1,\ldots,\breve{\psi}_{N-n},\ldots,\psi_N](x)}
  {\text{W}[\psi_1,\ldots,\psi_{N}](x)}
  =\frac{2^{n-N}(N-n)}{N!}\cdot(\cosh x)^{n-N}P^{(N-n,N-n)}_n(\tanh x),
\end{align}
and the scattering amplitudes
\begin{equation}
  \mathfrak{t}^{[N]}(k)=\prod_{j=1}^N\frac{(k+i\,j)}{(k-i\,j)},\quad
  \mathfrak{r}^{[N]}(k)=0,
\end{equation}
which is the $\gamma\to0$ limit of \eqref{tkrk_idQM} with $k_j=j$
($j=1,\ldots,N$).

Likewise the Hamiltonian with the {\em generic coupling\/} $h$,
\begin{equation}
  \mathfrak{H}=p^2-\frac{h(h+1)}{\cosh^2x}
  \ \ (-\infty<x<\infty),\quad h>0,
  \label{Hsol}
\end{equation}
has the eigenvalue $\mathcal{E}_n$ and the corresponding eigenfunction
$\phi_n(x)$ (for example see \cite{os29})
\begin{align}
  &\mathcal{E}_n=-(h-n)^2,\quad
  \phi_n(x)=(\cosh x)^{n-h}P_n^{(h-n,h-n)}(\tanh x)\ \ (n=0,1,\ldots,[h]'),
  \label{solitoneigen_oQM}
\end{align}
where $[x]'$ denotes the greatest integer not exceeding and not equal to $x$.
The transmission and the reflection amplitudes \cite{KS,HLS} are:
\begin{equation}
  \mathfrak{t}(k)=\frac{\Gamma(-h-ik)\Gamma(1+h-ik)}{\Gamma(-ik)\Gamma(1-ik)},
  \quad
  \mathfrak{r}(k)=\frac{\Gamma(ik)\Gamma(-h-ik)\Gamma(1+h-ik)}
  {\Gamma(-ik)\Gamma(-h)\Gamma(1+h)}.
  \label{soltr}
\end{equation}
The pole of the Gamma function $\Gamma(-h)$ in the denominator of
$\mathfrak{r}(k)$ gives the reflectionless potential $\mathfrak{r}(k)=0$
at $h=N\in\mathbb{Z}_{>0}$.
It is elementary to verify the unitarity relation
\begin{equation}
  |\mathfrak{t}(k)|^2+|\mathfrak{r}(k)|^2=1\ \ (k\in\mathbb{R}_{>0}).
  \label{solunit}
\end{equation}

For deriving the scattering amplitudes of this potential, the Jacobi
(Gegenbauer) polynomial in the eigenfunction \eqref{solitoneigen_oQM} is
replaced by the Gaussian hypergeometric function
\begin{equation*}
  (2\cosh x)^{n-h}{}_2F_1\Bigl(\genfrac{}{}{0pt}{}{-n,\,-n+2h+1}{h-n+1}
  \!\!\Bigm|\!\frac{1-\tanh x}{2}\Bigr).
\end{equation*}
Then it is analytically continued in the complex $k$ plane by the
substitution $n=h+ik$:
\begin{equation}
  (2\cosh x)^{ik}{}_2F_1\Bigl(\genfrac{}{}{0pt}{}{-h-ik,\,1+h-ik}{1-ik}
  \!\!\Bigm|\!\frac{1-\tanh x}{2}\Bigr),
  \label{unitwave}
\end{equation}
which goes asymptotically to the unit amplitude right moving wave $e^{ikx}$
at $x=+\infty$.
By using the {\em connection formula of the Gaussian hypergeometric function\/},
\begin{align}
  {}_2F_1\Bigl(\genfrac{}{}{0pt}{}{\alpha,\,\beta}{\gamma}\!\!\Bigm|\!z\Bigr)
  &=\phantom{+}\frac{\Gamma(\gamma)\Gamma(\alpha+\beta-\gamma)}
  {\Gamma(\alpha)\Gamma(\beta)}(1-z)^{\gamma-\alpha-\beta}\cdot
  {}_2F_1\Bigl(\genfrac{}{}{0pt}{}{\gamma-\alpha,\,\gamma-\beta}
  {\gamma-\alpha-\beta+1}\!\!\Bigm|\!{1-z}\Bigr)\n
  &\quad+\frac{\Gamma(\gamma)\Gamma(\gamma-\alpha-\beta)}
  {\Gamma(\gamma-\alpha)\Gamma(\gamma-\beta)}\cdot
  {}_2F_1\Bigl(\genfrac{}{}{0pt}{}{\alpha,\,\beta}
  {\alpha+\beta-\gamma+1}\!\!\Bigm|\!{1-z}\Bigr),
  \label{2F1_conn}
\end{align}
the asymptotic behaviour of the above wave function \eqref{unitwave} at
$x=-\infty$ is given by
\begin{equation}
  \frac{\Gamma(1-ik)\Gamma(-ik)}{\Gamma(-h-ik)\Gamma(1+h-ik)}\,e^{ikx}+
  \frac{\Gamma(1-ik)\Gamma(ik)}{\Gamma(1+h)\Gamma(-h)}\,e^{-ikx}.
\end{equation}
The unit amplitude right moving wave $e^{ikx}$ at $x=-\infty$ propagates
to $\mathfrak{t}(k)e^{ikx}$ at $x=+\infty$ and reflected to the left moving
wave $\mathfrak{r}(k)e^{-ikx}$ at $x=-\infty$, with the transmission amplitude
$\mathfrak{t}(k)$ and the reflection amplitude $\mathfrak{r}(k)$ as given
above \eqref{soltr}.

\subsection{Reflectionless case}
\label{sec:disreflesssoliton}

Next, the discrete QM counterpart goes as follows.
For the choice of $k_j$ and $\tilde{c}_j$ in \eqref{solitonkjctj}, the
Casoratian of $\psi_1,\ldots,\psi_N$ \eqref{psij_idQM} becomes
\begin{equation}
  \text{W}_{\gamma}[\psi_1,\ldots,\psi_N](x)
  =\prod_{j=1}^{N-1}\bigl(2\sin\tfrac{\gamma}{2}j\bigr)^{N-j}\cdot
  \prod_{j=1}^N\prod_{l=1}^j2\cosh\bigl(x+i\gamma(\tfrac{j+1}{2}-l)\bigr).
\end{equation}
By using this we obtain a very simple form of the potential function
$V^{[N]}(x)$ and the ground state wavefunction $\Phi^{[N]}_N(x)$:
\begin{align}
  &V^{[N]}(x)=e^{-i\gamma N}
  \frac{(1+e^{i\gamma N}e^{2x})(1+e^{i\gamma(N-1)}e^{2x})}
  {(1+e^{2x})(1+e^{-i\gamma}e^{2x})},
  \label{solitonVN}\\
  &\Phi^{[N]}_N(x)=\Bigl(\prod_{j=1}^{N-1}2\sin\tfrac{\gamma}{2}j\Bigr)^{-1}
  \cdot\Bigl(\prod_{j=1}^N4\cosh(x-i\tfrac{\gamma}{2}j)
  \cosh(x+i\tfrac{\gamma}{2}j)\Bigr)^{-\frac12}.
\end{align}
In the $\gamma\to 0$ limit, this Hamiltonian $\mathcal{H}^{[N]}$ with
$V^{[N]}(x)$ \eqref{solitonVN} reduces to the reflectionless $1/\cosh^2x$
potential $\mathfrak{U}^{[N]}(x)$ \eqref{solitonUN} in ordinary QM.
Moreover we have
\begin{align}
  &\quad\frac{\text{W}_{\gamma}
  [\psi_1,\ldots,\breve{\psi}_{N-n},\ldots,\psi_N](x)}
  {\text{W}_{\gamma}[\psi_1,\ldots,\psi_{N-1}](x)}\times
  \prod_{l=1}^n\frac{2\sin\frac{\gamma}{2}l\,\sin\frac{\gamma}{2}(2N-2n+l)}
  {\sin\frac{\gamma}{2}(N-l)}\n
  &=e^{-i\gamma(N-\frac14(3n-1))n}
  p_n(i\sinh x;e^{i\frac{\gamma}{2}N},e^{i\frac{\gamma}{2}(N-1)},
  -e^{i\frac{\gamma}{2}N},-e^{i\frac{\gamma}{2}(N-1)}|e^{-i\gamma}),
  \label{pnHN}
\end{align}
where $p_n(\eta;a_1,a_2,a_3,a_4|q)$ is the Askey-Wilson polynomial
\cite{koeswart}, expressed in terms of the basic hypergeometric function
${}_4\phi_3$:
\begin{align} 
  &\quad p_n(\cos x\,;a_1,a_2,a_3,a_4|q)
  \qquad(b_4\eqdef a_1a_2a_3a_4)\n
  &\eqdef a_1^{-n}(a_1a_2,a_1a_3,a_1a_4\,;q)_n\,
  {}_4\phi_3\Bigl(\genfrac{}{}{0pt}{}{q^{-n},\,b_4q^{n-1},\,
  a_1e^{ix},\,a_1e^{-ix}}{a_1a_2,\,a_1a_3,\,a_1a_4}\!\!\Bigm|\!q\,;q\Bigr).
  \label{defAW}
\end{align}
The type of parameter restrictions of the Askey-Wilson polynomial in
\eqref{pnHN} is also called the ({\em continuous\/}) {\em $q$-ultraspherical
polynomial} \cite{koeswart,askey}, which is a $q$-analogue of the Gegenbauer
polynomial. Because of the symmetry of the parameters, the $q$-ultraspherical
polynomial can also be expressed by ${}_3\phi_2$ or ${}_2\phi_1$
\cite{koeswart,askey}:
\begin{align}
  C_n(\cos x;\beta|q)
  &=\frac{(\beta^2;q)_n}{(\beta q^{\frac12},-\beta,-\beta q^{\frac12},q;q)_n}
  p_n(\cos x;\beta^{\frac12},\beta^{\frac12}q^{\frac12},-\beta^{\frac12},
  -\beta^{\frac12}q^{\frac12}|q)\n
  &=\frac{(\beta^2;q)_n}{(q;q)_n}\beta^{-\frac12n}
  {}_4\phi_3\Bigl(\genfrac{}{}{0pt}{}
  {q^{-n},\,\beta^2q^n,\,\beta^{\frac12}e^{ix},\,\beta^{\frac12}e^{-ix}}
  {\beta q^{\frac12},\,-\beta,\,-\beta q^{\frac12}}\!\!\Bigm|\!q\,;q\Bigr)\n
  &=\frac{(\beta^2;q)_n}{(q;q)_n}\beta^{-n}e^{-inx}
  {}_3\phi_2\Bigl(\genfrac{}{}{0pt}{}
  {q^{-n},\,\beta,\,\beta e^{2ix}}
  {\beta^2,\,0}\!\!\Bigm|\!q\,;q\Bigr)\n
  &=\frac{(\beta;q)_n}{(q;q)_n}e^{inx}
  {}_2\phi_1\Bigl(\genfrac{}{}{0pt}{}
  {q^{-n},\,\beta}
  {\beta^{-1}q^{1-n}}\!\!\Bigm|\!q\,;\beta^{-1}qe^{-2ix}\Bigr)
  \quad(n\in\mathbb{Z}_{\geq 0}).
  \label{quldef}
\end{align}
The change of the sinusoidal coordinates $\cos x\to i\sinh x$ is realised by
((3.39) in \cite{os32})
\begin{equation}
  \cos x\to i\sinh x\ \Longleftrightarrow\ x\to\tfrac{\pi}{2}-ix.
  \label{xchange}
\end{equation}
We choose the proportionality constant in the eigenfunction $\phi^{[N]}_n(x)$
of the reflectionless potential \eqref{phiN=PhiN_2} as
\begin{equation}
  \phi^{[N]}_n(x)=\prod_{l=1}^{N-1}2\sin\tfrac{\gamma}{2}l\cdot
  \prod_{l=1}^n\frac{2\sin\frac{\gamma}{2}l\,\sin\frac{\gamma}{2}(2N-2n+l)}
  {\sin\frac{\gamma}{2}(N-l)}
  \times\Phi^{[N]}_{N-n}(x).
  \label{solitonphiNj_idQM}
\end{equation}

\bigskip
In \cite{os32} we presented the exactly solvable discrete QM system with
$|q|=1$ and the sinusoidal coordinate $\eta(x)=\sinh x$.
Its Hamiltonian (with $K=1$) is
\begin{align}
  &\mathcal{H}'=\mathcal{A}^{\dagger}\mathcal{A},\quad
  \mathcal{A}\eqdef i\bigl(e^{\frac{\gamma}{2}p}\sqrt{V^*(x)}
  -e^{-\frac{\gamma}{2}p}\sqrt{V(x)}\,\bigr),
  \label{Hprime}\\
  &V(x)=e^{i\pi }e^{-i\frac{\gamma}{2}}\frac{a_1^*a_2^*}{|a_1a_2|}
  \frac{\prod_{j=1}^2(1+a_je^x)(1-a_j^{*\,-1}e^x)}
  {(1+e^{2x})(1+e^{-i\gamma}e^{2x})},\quad q=e^{-i\gamma}.
  \label{V(viii)}
\end{align}
For the parameters
\begin{equation}
  a_j=e^{-i\gamma(\alpha_j+i\beta_j)},
  \ \ \beta_j\in\mathbb{R},
  \ \  \gamma-\pi<\gamma\alpha_j<0,
  \ \ -\gamma\alpha>\pi-\tfrac{\gamma}{2}
  \ \ (\alpha\eqdef\alpha_1+\alpha_2),
  \label{pare(viii)}
\end{equation}
its eigenfunctions are
\begin{align}
  &\mathcal{H}'\phi_n(x)=\mathcal{E}'_n\phi_n(x)
  \ \ (n=0,1,\ldots,n_{\text{max}}),
  \ \ n_{\text{max}}=[\tfrac12-\alpha-\tfrac{\pi}{\gamma}]',\n
  &\phi_n(x)=\phi_0(x)P_n\bigl(\eta(x)\bigr),
  \ \ \eta(x)=\sinh x,
  \ \ \mathcal{E}'_n=
  4\sin\tfrac{\gamma}{2}n\,\sin\tfrac{\gamma}{2}(n-1+2\alpha),\n
  &P_n(\eta)=e^{-i\frac{\pi}{2}n}e^{i\gamma\frac34n(n-1)}e^{i\gamma\alpha n}\,
  (-i)^np_n(i\eta;ia_1,ia_2,-ia_1^{*\,-1},-ia_2^{*\,-1}|e^{-i\gamma}),
  \label{finPn}\\
  &\phi_0(x)=e^{(\frac12-\alpha-\frac{\pi}{\gamma})x}\sqrt{1+e^{2x}}\n
  &\phantom{\phi_0(x)=}\times\biggl(\prod_{j=1}^2\frac{
  \Phi_{\frac{\gamma}{2}}\bigl(x+\gamma\beta_j+i\gamma(\frac12-\alpha_j)\bigr)
  \Phi_{\frac{\gamma}{2}}\bigl(x-\gamma\beta_j+i\gamma(\frac12-\alpha_j)-i\pi
  \bigr)}
  {\Phi_{\frac{\gamma}{2}}\bigl(x+\gamma\beta_j-i\gamma(\frac12-\alpha_j)\bigr)
  \Phi_{\frac{\gamma}{2}}\bigl(x-\gamma\beta_j-i\gamma(\frac12-\alpha_j)+i\pi
  \bigr)}
  \biggr)^{\frac12}.
  \label{phi0(viii)}
\end{align}
{\em The quantum dilogarithm function\/} $\Phi_{\gamma}(z)$ \cite{qdilog}
plays the main role in the orthogonality weight function for the finitely many
orthogonal polynomials $P_n(\eta)$ \eqref{finPn} with $|q|=1$.
It is a meromorphic function, defined for $|\text{Im}\,z|<\gamma+\pi$ by
the integral representation
\begin{equation}
  \Phi_{\gamma}(z)=\exp\Bigl(\int_{\mathbb{R}+i0}
  \frac{e^{-izt}}{4\sinh\gamma t\,\sinh\pi t}\frac{dt}{t}\Bigr)
  \qquad(|\text{Im}\,z|<\gamma+\pi),
  \label{intrep}
\end{equation}
and analytically continued to the whole complex plane by the functional equation
\begin{equation}
  \frac{\Phi_{\gamma}(z+i\gamma)}{\Phi_{\gamma}(z-i\gamma)}
  =\frac{1}{1+e^z}.
  \label{funcrel}
\end{equation}
For more properties of the quantum dilogarithm functions, see Appendix B
in \cite{os32}.

Now let us identify the obtained potential function $V^{[N]}(x)$
\eqref{solitonVN} as a special case of the above exactly solvable $|q|=1$
systems.
We assume
\begin{equation}
  N+2<\frac{\pi}{\gamma},
\end{equation}
and choose the parameters of $\mathcal{H}'$ in \eqref{pare(viii)}
\begin{equation}
  \beta_1=\beta_2=0,
  \ \ \alpha_1=-\tfrac{\pi}{2\gamma}-\tfrac{N}{2},
  \ \ \alpha_2=-\tfrac{\pi}{2\gamma}-\tfrac{N-1}{2}
  \ \ \bigl(\Rightarrow a_1=ie^{i\frac{\gamma}{2}N},
  \ a_2=ie^{i\frac{\gamma}{2}(N-1)}\bigr).
  \label{para_relsol_idQM}
\end{equation}
Then the two Hamiltonian $\mathcal{H}^{[N]}$ with $V^{[N]}(x)$
\eqref{solitonVN} and $\mathcal{H}'$ \eqref{Hprime} are equal up to an
additive constant
\begin{equation}
  \mathcal{H}^{[N]}=\mathcal{H}'+\tilde{\mathcal{E}}_{k_N},
  \label{HNHprimeeq}
\end{equation}
and the ground state wave function $\phi_0(x)$ \eqref{phi0(viii)} and the
eigenpolynomials $P_n(\eta)$ \eqref{finPn} of $\mathcal{H}'$ become
\begin{align}
  &\phi_0(x)=\Bigl(\prod_{j=1}^N4\cosh(x-i\tfrac{\gamma}{2}j)
  \cosh(x+i\tfrac{\gamma}{2}j)\Bigr)^{-\frac12},
  \label{phi0_N_idQM}\\
  &P_n(\eta)=e^{-i\gamma(N-\frac14(3n-1))n}
  p_n(i\eta;e^{i\frac{\gamma}{2}N},e^{i\frac{\gamma}{2}(N-1)},
  -e^{i\frac{\gamma}{2}N},-e^{i\frac{\gamma}{2}(N-1)}|e^{-i\gamma}).
\end{align}
Namely $\phi_n(x)$ \eqref{phi0(viii)} of $\mathcal{H}'$ agrees with
$\phi^{[N]}_n(x)$ \eqref{solitonphiNj_idQM} of $\mathcal{H}^{[N]}$.
The additive constant $\tilde{\mathcal{E}}_{k_N}$ in \eqref{HNHprimeeq}
provides the proper eigenvalues of $\mathcal{H}^{[N]}$
\begin{equation*}
  \mathcal{E}^{[N]}_n
  =\mathcal{E}'_n+\tilde{\mathcal{E}}_{k_N}
  =4\sin\tfrac{\gamma}{2}n\,\sin\tfrac{\gamma}{2}(n-1+2\alpha)
  -4\sin^2\tfrac{\gamma}{2}k_N
  =-4\sin^2\tfrac{\gamma}{2}(N-n)=\tilde{\mathcal{E}}_{k_{N-n}},
\end{equation*}
and the highest level of discrete eigenstates is also
$n_{\text{max}}=[\frac12-\alpha-\frac{\pi}{\gamma}]'=[N]'=N-1$.
In order to show that the quantum dilogarithm in the ground state wave
function $\phi_0(x)$ \eqref{phi0(viii)} of $\mathcal{H}'$ reduces to the
elementary function \eqref{phi0_N_idQM}, the functional equation
\eqref{funcrel} is used repeatedly,
\begin{equation}
  \frac{\Phi_{\gamma}(z+i\gamma n_1)}{\Phi_{\gamma}(z-i\gamma n_2)}
  =\prod_{k=0}^{\frac{n_1+n_2}{2}-1}\!\!
  \frac{1}{1+e^{z+i\gamma(n_1-2k-1)}}\quad
  \bigl(n_1,n_2\in\mathbb{Z}_{\geq 0},\ \ n_1\equiv n_2\ (\text{mod $2$})\bigr).
\end{equation}
Thus the discrete analogue of the {\em reflectionless\/} $1/\cosh^2x$
potential is constructed by the special combination of the discrete
reflectionless potential.

\subsection{Generic case}
\label{sec:disgensoliton}

Now we turn to the construction of the discrete analogue of {\em generic\/}
$1/\cosh^2x$ potential.
Let us consider $\mathcal{H}'$ \eqref{Hprime} with the parameters,
\begin{align}
  &\beta_1=\beta_2=0,
  \ \ \alpha_1=-\tfrac{\pi}{2\gamma}-\tfrac{h}{2},
  \ \ \alpha_2=-\tfrac{\pi}{2\gamma}-\tfrac{h}{2}+\tfrac12
  \ \ \bigl(\Rightarrow a_1=ie^{i\frac{\gamma}{2}h},
  \ a_2=ie^{i\frac{\gamma}{2}(h-1)}\bigr),\n
  &h>0,\quad h+2<\tfrac{\pi}{\gamma},
  \label{para_sol_idQM}
\end{align}
and define a new Hamiltonian $\mathcal{H}$:
\begin{equation}
  \mathcal{H}\eqdef\mathcal{H}'+\tilde{\mathcal{E}}_h.
  \label{genham}
\end{equation}
The parameter ranges satisfy the previous restrictions \eqref{pare(viii)}.
The eigenstates of $\mathcal{H}$ are
\begin{align}
  &\mathcal{H}\phi_n(x)=\mathcal{E}_n\phi_n(x)
  \ \ (n=0,1,\ldots,n_{\text{max}}),
  \ \ n_{\text{max}}=[h]',\n
  &V(x)=e^{-i\gamma h}\frac{(1+e^{i\gamma h}e^{2x})
  (1+e^{i\gamma(h-1)}e^{2x})}{(1+e^{2x})(1+e^{-i\gamma}e^{2x})},
  \label{Vgen}\\
  &\phi_n(x)=\phi_0(x)P_n\bigl(\eta(x)\bigr),
  \ \ \eta(x)=\sinh x,
  \ \ \mathcal{E}_n=\tilde{\mathcal{E}}_{h-n}=-4\sin^2\tfrac{\gamma}{2}(h-n),
  \label{phigen}\\
  &P_n(\eta)=e^{-i\gamma(h-\frac14(3n-1))n}
  p_n(i\eta;e^{i\frac{\gamma}{2}h},e^{i\frac{\gamma}{2}(h-1)},
  -e^{i\frac{\gamma}{2}h},-e^{i\frac{\gamma}{2}(h-1)}|e^{-i\gamma})\n
  &\phantom{P_n(\eta)}\propto C_n(i\sinh x;\beta|q),\quad q=e^{-i\gamma},
  \quad \beta\eqdef e^{i\gamma h}=q^{-h},
  \label{polygen}\\
  &\phi_0(x)=e^{hx}\sqrt{1+e^{2x}}\n
  &\phantom{\phi_0(x)=}\times\biggl(\frac{
  \Phi_{\frac{\gamma}{2}}\bigl(x+i\frac{\gamma}{2}(h+1)+i\frac{\pi}{2}\bigr)
  \Phi_{\frac{\gamma}{2}}\bigl(x+i\frac{\gamma}{2}(h+1)-i\frac{\pi}{2}
  \bigr)}
  {\Phi_{\frac{\gamma}{2}}\bigl(x-i\frac{\gamma}{2}(h+1)-i\frac{\pi}{2}\bigr)
  \Phi_{\frac{\gamma}{2}}\bigl(x-i\frac{\gamma}{2}(h+1)+i\frac{\pi}{2}
  \bigr)}\n
  &\phantom{\phi_0(x)=\times}\times\frac{
  \Phi_{\frac{\gamma}{2}}\bigl(x+i\frac{\gamma}{2}h+i\frac{\pi}{2}\bigr)
  \Phi_{\frac{\gamma}{2}}\bigl(x+i\frac{\gamma}{2}h-i\frac{\pi}{2}
  \bigr)}
  {\Phi_{\frac{\gamma}{2}}\bigl(x-i\frac{\gamma}{2}h-i\frac{\pi}{2}\bigr)
  \Phi_{\frac{\gamma}{2}}\bigl(x-i\frac{\gamma}{2}h+i\frac{\pi}{2}
  \bigr)}
  \biggr)^{\frac12}\n
  &\phantom{\phi_0(x)}=e^{hx}\sqrt{1+e^{2x}}
  \ \biggl(\frac{
  \Phi_{\gamma}\bigl(2x+i\gamma(1+h)\bigr)
  \Phi_{\gamma}\bigl(2x+i\gamma h\bigr)}
  {\Phi_{\gamma}\bigl(2x-i\gamma(1+h)\bigr)
  \Phi_{\gamma}\bigl(2x-i\gamma h\bigr)}\biggr)^{\frac12}\n
  &\phantom{\phi_0(x)}=e^{hx}\sqrt{1+e^{2x}}
  \ \biggl(\frac{
  \Phi_{\frac{\gamma}{2}}\bigl(2x+i\gamma(h+\frac12)\bigr)}
  {\Phi_{\frac{\gamma}{2}}\bigl(2x-i\gamma(h+\frac12)\bigr)}
  \biggr)^{\frac12},
  \label{qultran}
\end{align}
where we have used the properties
\begin{equation}
  \Phi_{\gamma}(z+i\tfrac{\pi}{2})\Phi_{\gamma}(z-i\tfrac{\pi}{2})
  =\Phi_{2\gamma}(2z),\quad 
    \Phi_{\gamma}(z+i\tfrac{\gamma}{2})\Phi_{\gamma}(z-i\tfrac{\gamma}{2})
  =\Phi_{\frac{\gamma}{2}}(z).
  \label{funcrel2}
\end{equation}
Note that $P_n(\eta)$ has definite parity $P_n(-\eta)=(-1)^nP_n(\eta)$, like
the Gegenbauer polynomial $P_n$ in \eqref{solitoneigen_oQM}.
It should be stressed that this Hamiltonian $\mathcal{H}$ \eqref{genham} is
symmetric under the parameter inversion:
\begin{align}
  h+1&\leftrightarrow-h,
  \label{parainv}\\
  V(x)V^*(x-i\gamma),&\quad
  V(x)+V^*(x)-\tilde{\mathcal{E}}_h:\ \text{invariant}.
\end{align}
Note that the eigenfunctions \eqref{phigen}--\eqref{qultran}
are not invariant under the parameter inversion because half of the solutions
are discarded to ensure the square integrability.
On the other hand, the scattering amplitudes are invariant since the full
two-dimensional solution space is needed for the scattering problem.
This Hamiltonian $\mathcal{H}$ has the correct $\gamma\to 0$ limit \cite{os32},
\begin{equation}
  \lim_{\gamma\to 0}\gamma^{-2}\mathcal{H}=\mathfrak{H}
  =p^2-\frac{h(h+1)}{\cosh^2x},
\end{equation}
which has {\em regular singular points} at $x=i\pi/2$, mod $i\pi$, with
the characteristic exponents $h+1$ and $-h$.
The invariance under the above parameter inversion \eqref{parainv} is
inherited by the Hamiltonian and by the {\em transmission and reflection
amplitudes\/} $\mathfrak{t}(k)$, $\mathfrak{r}(k)$ \eqref{soltr}.

\subsection{Scattering amplitudes: Conjectures}
\label{sec:conject}

In ordinary QM, scattering problems can be formulated for non-confining
potentials, that is, those having finitely many discrete eigenstates defined
on a full line or a half line. They are {\em exactly solvable, i.e. the
transmission and reflection amplitudes are exactly calculable, if the discrete
eigenvalues and eigenfunctions are exactly known\/} \cite{KS,HLS}.
We naturally expect the same situation in discrete QM and the scattering
problem for the Hamiltonian $\mathcal{H}$ \eqref{genham} should also be
exactly solvable.
Asymptotically the Hamiltonian $\mathcal{H}$ \eqref{genham} has the plane
wave solutions $e^{\pm ikx}$, since it approaches to the free Hamiltonian
$\mathcal{H}_0$ \eqref{H0_idQM}:
\begin{equation*}
  \mathcal{H}\to\mathcal{H}_0\ \ (x\to\pm\infty),\quad
  \mathcal{H}e^{\pm ikx}\approx\mathcal{E}_k^{\text{s}}e^{\pm ikx},\quad
  \mathcal{E}_k^{\text{s}}=4\sinh^2\frac{k\gamma}2.
\end{equation*}

The scattering problem is formulated as follows.
We pick up a special wave solution $\Psi_k(x)$ which approaches to the unit
amplitude right moving plane wave at $x\to+\infty$:
\begin{equation}
  \mathcal{H}\Psi_k(x)=\mathcal{E}_k^{\text{s}}\Psi_k(x),\qquad
  \Psi_k(x)\to e^{ikx}\ \ (x\to+\infty).
\end{equation}
Analytically continued in $x$ to the region $x=-\infty$, $\Psi_k(x)$ is a
linear combination of the right and left moving plane waves:
\begin{equation}
  \Psi_k(x)\to A(k)e^{ikx}+B(k)e^{-ikx}\ \ (x\to-\infty).
\end{equation}
The transmission $t(k)$ and the reflection $r(k)$ amplitudes are defined by
\begin{equation}
  t(k)\eqdef\frac{1}{A(k)},\quad r(k)\eqdef\frac{B(k)}{A(k)}.
\end{equation}
At a possible zero of $A(k)$ (a pole of $t(k)$) on the positive imaginary
$k$ axis, $k=i\kappa$ ($\kappa>0$):
\begin{equation*}
  A(i\kappa)=0,
\end{equation*}
the special wave solution $\Psi_k(x)$ analytically continued in the complex
upper $k$ plane has the asymptotic behaviours of a bound state 
\begin{equation}
  \Psi_{i\kappa}(x)\to\left\{
  \begin{array}{rl}
  e^{-\kappa x}&:x\to+\infty\\
  B(i\kappa)e^{\kappa x}&:x\to-\infty 
  \end{array}\right.,
\end{equation}
with the eigenvalue
$\tilde{\mathcal E}_{\kappa}(=\mathcal{E}^{\text{s}}_{i\kappa})$.

The special plane wave solution is obtained from the discrete eigenfunction
\eqref{polygen}--\eqref{qultran} expressed by the basic hypergeometric function:
\begin{align*}
  \phi_n(x)&\propto\phi_0(x)C_n(i\sinh x;\beta|q),\qquad q=e^{-i\gamma},\quad
  \beta=e^{i\gamma h}=q^{-h},\n
  &\propto e^{hx}\sqrt{1+e^{2x}}
  \ \biggl(\frac{
  \Phi_{\frac{\gamma}{2}}\bigl(2x+i\gamma(h+\frac12)\bigr)}
  {\Phi_{\frac{\gamma}{2}}\bigl(2x-i\gamma(h+\frac12)\bigr)}
  \biggr)^{\frac12} e^{nx} {}_2\phi_1\Bigl(\genfrac{}{}{0pt}{}
  {q^{-n},\,\beta}
  {\beta^{-1}q^{1-n}}\!\!\Bigm|\!q\,;\beta^{-1}qe^{-2x-i\pi}\Bigr),
\end{align*}
where the substitution of the coordinate $x\to\tfrac{\pi}2-ix$ \eqref{xchange}
is made corresponding to the sinusoidal coordinates change $\cos x\to i\sinh x$
in the $q$-ultraspherical polynomial \eqref{quldef}.
Next it is analytically continued into the complex $k$ plane by the
substitution $n= h+ik$ as in the ordinary QM:
\begin{equation}
  \Psi_k(x)\eqdef e^{ikx}e^{2hx}\sqrt{1+e^{2x}}\ \biggl(\frac{
  \Phi_{\frac{\gamma}{2}}\bigl(2x+i\gamma(h+\frac12)\bigr)}
  {\Phi_{\frac{\gamma}{2}}\bigl(2x-i\gamma(h+\frac12)\bigr)}
  \biggr)^{\frac12}{}_2\phi_1\Bigl(\genfrac{}{}{0pt}{}
  {q^{-h-ik},\,q^{-h}}
  {q^{1-ik}}\!\!\Bigm|\!q\,;q^{1+h}e^{-2x-i\pi}\Bigr),
  \label{unitr}
\end{equation}
which approaches to the unit amplitude right moving wave $e^{ikx}$ at
$x\to \infty$.
Hereafter we exclude the {\em cases of $q$ being a root of unity $q^n=1$
$(n\in\mathbb{Z}_{>0})$\/}.
The convergence of the basic hypergeometric function
${}_2\phi_1(\genfrac{}{}{0pt}{}{a,\,b}{c}|q\,;z)$ with $|q|=1$ is a subtle
problem and we do not know if the answer is known.
We consider $|q|=1$ is approached from below, $|q|\nearrow1$ \cite{aaa},
which is realised by adding an infinitesimal negative imaginary part to
$\gamma$, $\gamma\to\gamma-i\epsilon$ ($\epsilon>0$).
In the following, we assume that the function
${}_2\phi_1(\genfrac{}{}{0pt}{}{a,\,b}{c}|q\,;z)$ with $|q|=1$ has a
positive radius of convergence at $z=0$ and it is analytically continued
to the whole complex $z$ plane.

\bigskip
In order to evaluate $\Psi_k(x)$ in the asymptotic region $x\to-\infty$,
{\em we need the connection formula for the basic hypergeometric function 
${}_2\phi_1(\genfrac{}{}{0pt}{}{a,\,b}{c}|q\,;z)$ with $|q|=1$\/}.
However, the authors are aware of the connection formula {\em only for\/}
$0<q<1$ (4.3.2) in \cite{gasper}:
\begin{align}
  {}_2\phi_1\Bigl(\genfrac{}{}{0pt}{}{a,\,b}{c}\!\!\Bigm|\!q\,;z\Bigr)
  &=\ \ \frac{(b,c/a;q)_\infty}{(c,b/a;q)_\infty}\,
  \frac{(az,q/(az);q)_\infty}{(z,q/z;q)_\infty}\cdot
  {}_2\phi_1\Bigl(\genfrac{}{}{0pt}{}{a,\,aq/c}
  {aq/b}\!\!\Bigm|\!q\,;\frac{cq}{abz}\Bigr)\n
  &\quad
  +\frac{(a,c/b;q)_\infty}{(c,a/b;q)_\infty}\,
  \frac{(bz,q/(bz);q)_\infty}{(z,q/z;q)_\infty}\cdot
  {}_2\phi_1\Bigl(\genfrac{}{}{0pt}{}{b,\,bq/c}
  {bq/a}\!\!\Bigm|\!q\,;\frac{cq}{abz}\Bigr),
  \label{watform}
\end{align}
derived by Watson \cite{wat}.
The underlying logic for the formula is that the basic hypergeometric
functions on the right hand side with the $z$-depending factors satisfy 
the same difference equation as the left hand side:
\begin{align}
  \Bigl((c -abz)q^{D_z}+(a+b)z-c-q+(q-z)q^{-D_z}\Bigr)\,
  {}_2\phi_1\Bigl(\genfrac{}{}{0pt}{}{a,\,b}{c}\!\!\Bigm|\!q\,;z\Bigr)=0,
  \label{q-difeq}
\end{align}
in which $D_z$ is $D_z\eqdef z\frac{d}{dz}$ and $q^{D_z}f(z)=f(qz)$.
This situation is unchanged when the base is changed from $0<q<1$ to $|q|=1$.
For $|q|=1$, however, the infinite $q$-shifted factorial $(a;q)_\infty$ does
not converge. We know through the construction of exactly solvable discrete
QM systems with $|q|=1$ \cite{os32}, that the infinite $q$-shifted factorials
should be replaced by quantum dilogarithm functions \eqref{intrep}.
Experience with $q$-orthogonal polynomials with $|q|=1$ \cite{os32} has led
us to the following replacement rule for the coefficients of the connection
formula 
\begin{equation}
  (e^z;q)_\infty\ \longrightarrow
  {\rm const.}/\Phi_{\frac{\gamma}{2}}^{(+)}(z),\qquad
  \Phi_{\frac{\gamma}{2}}^{(+)}(z)\eqdef
  \Phi_{\frac{\gamma}{2}}\bigl(z+i\tfrac{\gamma}{2}+i\pi\bigr),\quad
   q=e^{-i\gamma}.
  \label{trrule}
\end{equation}
The above constant factor is not yet determined, since the infinite
$q$-factorials appear in ratios.
By applying this replacement rule to the connection formula
with $0<q<1$ above \eqref{watform}, we arrive at the following:
\begin{conj}\label{conju1}
The connection formula for the basic hypergeometric function ${}_2\phi_1$
with $|q|=1$, $(q=e^{-i\gamma},\,\gamma>0)$ reads:
\begin{align}
  &\quad{}_2\phi_1\Bigl(\genfrac{}{}{0pt}{}{e^{\lambda},\,e^{\mu}}
  {e^{\nu}}\!\!\Bigm|\!q\,;e^{z}\Bigr)
  \label{phicon}\\
  &=\phantom{+}
  \frac{\Phi_{\frac{\gamma}{2}}^{(+)}({\nu})
  \Phi_{\frac{\gamma}{2}}^{(+)}(\mu-\lambda)}
  {\Phi_{\frac{\gamma}{2}}^{(+)}({\mu})
  \Phi_{\frac{\gamma}{2}}^{(+)}({\nu-\lambda})}\,
  \frac{\Phi_{\frac{\gamma}{2}}^{(+)}({z})
  \Phi_{\frac{\gamma}{2}}^{(+)}({-i\gamma-z})}
  {\Phi_{\frac{\gamma}{2}}^{(+)}({\lambda+z})
  \Phi_{\frac{\gamma}{2}}^{(+)}({-i\gamma-\lambda-z})}
  \cdot{}_2\phi_1\Bigl(\genfrac{}{}{0pt}{}
  {e^{\lambda},\,q\,e^{\lambda-\nu}}
  {q\,e^{\lambda-\mu}}\!\!\Bigm|\!q\,;q\,e^{\nu-\lambda-\mu-z}\Bigr)\n[2pt]
  &\quad+
  \frac{\Phi_{\frac{\gamma}{2}}^{(+)}({\nu})
  \Phi_{\frac{\gamma}{2}}^{(+)}(\lambda-\mu)}
  {\Phi_{\frac{\gamma}{2}}^{(+)}({\lambda})
  \Phi_{\frac{\gamma}{2}}^{(+)}({\nu-\mu})}\,
  \frac{\Phi_{\frac{\gamma}{2}}^{(+)}({z})
  \Phi_{\frac{\gamma}{2}}^{(+)}({-i\gamma-z})}
  {\Phi_{\frac{\gamma}{2}}^{(+)}({\mu+z})
  \Phi_{\frac{\gamma}{2}}^{(+)}({-i\gamma-\mu-z})}
  \cdot{}_2\phi_1\Bigl(\genfrac{}{}{0pt}{}
  {e^{\mu},\,q\,e^{\mu-\nu}}
  {q\,e^{\mu-\lambda}}\!\!\Bigm|\!q\,;q\,e^{\nu-\lambda-\mu-z}\Bigr).
  \nonumber
\end{align}
\end{conj}
\begin{rema}\label{rem:1}
The first factors of the above connection formula and the arguments of
${}_2\phi_1$ have similar structures to those of the corresponding Gaussian
hypergeometric function:
\begin{align}
  {}_2F_1\Bigl(\genfrac{}{}{0pt}{}{\lambda,\,\mu}{\nu}\!\!\Bigm|\!z\Bigr)
  &=\phantom{+}\frac{\Gamma(\nu)\Gamma(\mu-\lambda)}
  {\Gamma(\mu)\Gamma(\nu-\lambda)}(-z)^{-\lambda}\cdot
  {}_2F_1\Bigl(\genfrac{}{}{0pt}{}{\lambda,\,\lambda-\nu+1}
  {\lambda-\mu+1}\!\!\Bigm|\!\frac{1}{z}\Bigr)\n[2pt]
  &\quad+\frac{\Gamma(\nu)\Gamma(\lambda-\mu)}
  {\Gamma(\lambda)\Gamma(\nu-\mu)}(-z)^{-\mu}\cdot
  {}_2F_1\Bigl(\genfrac{}{}{0pt}{}{\mu,\,\mu-\nu+1}
  {\mu-\lambda+1}\!\!\Bigm|\!\frac{1}{z}\Bigr).
  \label{2F1_con}
\end{align}
\end{rema}
\begin{rema}\label{rem:0}
It is straightforward to show that both of the basic hypergeometric functions
with the $z$-dependent factors on the right hand side satisfy the same
$q$-difference equation \eqref{q-difeq} as the left hand side.
One only has to apply translation rules
\begin{equation*}
  e^{\lambda}\to a,\quad e^{\mu}\to b,\quad e^{\nu}\to c,\quad
  e^z\to z,\quad e^{-i\gamma}\to q,\quad\text{etc}.
\end{equation*}
The quantum dilogarithm formulas \eqref{funcrel},
\eqref{funcrel2}, \eqref{poles}--\eqref{asymp} are useful.
\end{rema}
\begin{rema}\label{rem:2}
Since we do not have an analytic proof for the connection formula
\eqref{phicon}, we will provide several supporting evidences for the
{\bf Conjecture\,\ref{conju1}} and
the replacement rule \eqref{trrule} in {\bf Appendix}.
\end{rema}

\bigskip
Based on the {\bf Conjecture\,\ref{conju1}}, the asymptotic form of the
plane wave solution $\Psi_k(x)$ at $x=-\infty$ is obtained:
\begin{align}
  \Psi_k(x)&\to e^{ikx}e^{-i\frac{\gamma}{2}h(h+1)}
  \frac{\Phi_{\frac{\gamma}{2}}^{(+)}(-k\gamma-i\gamma)
  \Phi_{\frac{\gamma}{2}}^{(+)}(-k\gamma)}
  {\Phi_{\frac{\gamma}{2}}^{(+)}(-k\gamma+i\gamma h)
  \Phi_{\frac{\gamma}{2}}^{(+)}(-k\gamma-i\gamma(h+1))}\n
  &\quad+e^{-ikx}e^{-i\frac{\gamma}{2}h(h+1)+\frac{k\gamma}{2}(1-ik)}
  \frac{\Phi_{\frac{\gamma}{2}}^{(+)}(-k\gamma-i\gamma)
  \Phi_{\frac{\gamma}{2}}^{(+)}(k\gamma)}
  {\Phi_{\frac{\gamma}{2}}^{(+)}(i\gamma h)
  \Phi_{\frac{\gamma}{2}}^{(+)}(-i\gamma(h+1))}.
\end{align}
This leads to the transmission and the reflection amplitudes for the
Hamiltonian $\mathcal{H}$ \eqref{genham}:
\begin{align}
  t(k)&=e^{i\frac{\gamma}{2}h(h+1)}
  \frac{\Phi_{\frac{\gamma}{2}}^{(+)}(-k\gamma+i\gamma h)
  \Phi_{\frac{\gamma}{2}}^{(+)}(-k\gamma-i\gamma(h+1))}
  {\Phi_{\frac{\gamma}{2}}^{(+)}(-k\gamma)
  \Phi_{\frac{\gamma}{2}}^{(+)}(-k\gamma-i\gamma)} \n
  &=e^{i\frac{\gamma}{2}h(h+1)}
  \frac{\Phi_{\frac{\gamma}{2}}(-k\gamma+i\gamma(h+\frac12)+i\pi)
  \Phi_{\frac{\gamma}{2}}(-k\gamma-i\gamma(h+\frac12)+i\pi)}
  {\Phi_{\frac{\gamma}{2}}(-k\gamma+i\frac{\gamma}{2}+i\pi)
  \Phi_{\frac{\gamma}{2}}(-k\gamma-i\frac{\gamma}{2}+i\pi)},
  \label{tgen}\\
  r(k)&=e^{\frac{k\gamma}{2}(1-ik)}
  \frac{\Phi_{\frac{\gamma}{2}}^{(+)}(k\gamma)
  \Phi_{\frac{\gamma}{2}}^{(+)}(-k\gamma+i\gamma h)
  \Phi_{\frac{\gamma}{2}}^{(+)}(-k\gamma-i\gamma(h+1))}
  {\Phi_{\frac{\gamma}{2}}^{(+)}(-k\gamma)
  \Phi_{\frac{\gamma}{2}}^{(+)}(i\gamma h)
  \Phi_{\frac{\gamma}{2}}^{(+)}(-i\gamma (h+1))}\n
  &=e^{\frac{k\gamma}{2}(1-ik)}
  \frac{\Phi_{\frac{\gamma}{2}}(k\gamma+i\frac{\gamma}{2}+i\pi)
  \Phi_{\frac{\gamma}{2}}(-k\gamma+i\gamma(h+\frac12)+i\pi)
  \Phi_{\frac{\gamma}{2}}(-k\gamma-i\gamma(h+\frac12)+i\pi)}
  {\Phi_{\frac{\gamma}{2}}(-k\gamma+i\frac{\gamma}{2}+i\pi)
  \Phi_{\frac{\gamma}{2}}(i\gamma(h+\frac12)+i\pi)
  \Phi_{\frac{\gamma}{2}}(-i\gamma (h+\frac12)+i\pi)}.
  \label{rgen}
\end{align}
They satisfy all the known criteria for the scattering amplitudes.
Obviously they are invariant under the parameter inversion
$-h\leftrightarrow h+1$ \eqref{parainv}.
The pole of the quantum dialog
$\Phi_{\frac{\gamma}{2}}\bigl(-k\gamma+i\gamma(h+\frac12)+i\pi\bigr)$
in the numerator of $t(k)$ ($k=i\kappa$, $\kappa>0$) at
\begin{equation}
  k=i(h-n),\quad n=0,1,\ldots,[h]',
\end{equation}
corresponds to the eigenvalue
$\mathcal{E}_n=\mathcal{E}^{\text{s}}_{i(h-n)}=-4\sin^2\frac{\gamma}{2}(h-n)$
in \eqref{phigen}.
For $0<\kappa\leq\frac{\pi}{\gamma}$, there are no other poles of $t(k)$.
The scattering amplitudes satisfy the unitarity relation
\begin{equation}
  |t(k)|^2+|r(k)|^2=1\ \ (k\in\mathbb{R}_{>0}).
\end{equation}
When $h$ is a positive integer $N$, they reduce to the special case obtained
in \S\,\ref{sec:refless} \eqref{tkrk_idQM} for $k_j=j$ ($j=1,\ldots,N$):
\begin{equation}
  t(k)=\prod_{j=1}^N\frac{\sinh\frac{\gamma}{2}(k+i\,j)}
  {\sinh\frac{\gamma}{2}(k-i\,j)},\quad r(k)=0.
\end{equation}
The pole of the quantum dialog
$\Phi_{\frac{\gamma}{2}}\bigl(i\gamma(h+\frac12)+i\pi\bigr)$ in the
denominator of $r(k)$ at $h=N$ is responsible for the reflectionless property.
There are no other zeros of $r(k)$.
Thus we do believe these scattering amplitudes are correct and they provide
a strong evidence for the connection formula, {\bf Conjecture\,\ref{conju1}}.
We do hope experts to provide an analytic proof of the connection formula.

For the verification of these results, on top of \eqref{funcrel},
\eqref{funcrel2}, the following properties and formulas are useful
\cite{os32}:\\
\underline{functional relations}:
\begin{align}
%
  &\frac{\Phi_{\gamma}(z+i\pi)}{\Phi_{\gamma}(z-i\pi)}
  =\frac{1}{1+e^{\frac{\pi}{\gamma}z}},\quad
  \Phi_{\gamma}(z)^*=\frac{1}{\Phi_{\gamma}(z^*)}
  \ (\text{complex conjugation}),
  \label{poles}\\
  &\Phi_{\gamma}(z)\Phi_{\gamma}(-z)=
  \exp\Bigl(\frac{i}{4\gamma}\bigl(z^2+\frac{\gamma^2+\pi^2}{3}\bigl)\Bigr).
\end{align}
\underline{poles and zeros}:
\begin{align}
  \text{poles of $\Phi_{\gamma}(z)$} &:
  z=i\bigl((2n_1-1)\gamma+(2n_2-1)\pi\bigr)\quad
  (n_1,n_2\in\mathbb{Z}_{\geq 1}),
  \label{pole}\\
  \text{zeros of $\Phi_{\gamma}(z)$} &:
  z=-i\bigl((2n_1-1)\gamma+(2n_2-1)\pi\bigr)\quad
  (n_1,n_2\in\mathbb{Z}_{\geq 1}).
  \label{zero}
\end{align}
\underline{asymptotic forms}: ($|\text{Im}\,z|<\gamma+\pi$)
\begin{equation}
  \Phi_{\gamma}(z)=
  \begin{cases}
  \exp\Bigl(\frac{i}{4\gamma}\bigl(z^2+\frac{\gamma^2+\pi^2}{3}\bigr)\Bigl)
  &\!\!(\text{Re}\,z\to\infty)\\[2pt]
  1&\!\!(\text{Re}\,z\to-\infty)
  \end{cases}.
 \label{asymp}
\end{equation}

\section{Summary and Comments}
\label{sec:summary}

In ordinary QM, the scattering problems for non-confining exactly solvable
systems defined on the full line or half line are known to be exactly solvable
\cite{KS,HLS}. We have recently developed several exactly solvable systems
with non-confining potentials in discrete QM with pure imaginary shifts
\cite{os32}. The eigenpolynomials are $q$-orthogonal polynomials with $|q|=1$
and their orthogonality weight functions are quantum dilogarithm functions.
One of them is the discrete counterpart of the $1/\cosh^2x$ potential.
We do expect that its scattering problem is exactly solvable.
However, the solutions of scattering problems in general require the
connection formulas of the (basic) hypergeometric functions.
To the best of our knowledge, the connection formula for ${}_2\phi_1$ with
$|q|=1$ is not known.
We made a conjecture of the connection formula {\bf Conjecture\,\ref{conju1}}
\eqref{phicon}, built upon the empirical correspondence between the infinite
$q$-shifted factorial and the quantum dilogarithm \eqref{trrule}.
Based on the conjectured connection formula for ${}_2\phi_1$ with $|q|=1$,
the scattering amplitudes of the discrete analogue of the $1/\cosh^2x$
potential are derived \eqref{tgen}--\eqref{rgen}. They reproduce well the
reflectionless limit and retain the proper symmetry of the Hamiltonian.
We do believe that the {\bf Conjecture\,\ref{conju1}} \eqref{phicon} is correct
and ask {\em experts to provide its analytical proof.\/}

It is also interesting to investigate the scattering problems of the exactly
solvable systems with non-confining potentials developed in \cite{os32};
the discrete counterpart of the Morse potential with the sinusoidal coordinate
$\eta(x)=e^{\pm x}$, the deformation of the hyperbolic
Darboux-P\"{o}schl-Teller potential with $\eta(x)=\cosh x$, the discrete
counterpart of the the hyperbolic symmetric top $\II$ with $\eta(x)=\sinh x$.

As for the discrete quantum mechanics with real shifts \cite{os12},
we have not yet developed a satisfactory scattering theory.

After completing the present work, several related papers were brought to
our attention. We thank friends and colleagues for the useful information.
The analytic difference operators having exponential interaction terms and
their eigenfunctions were discussed in \cite{rui}. The $q$-hypergeometric
function of the Barnes type with $|q|=1$ and its connection formula were
presented in \cite{nishue,take}. Quantum dilogarithm functions appeared in
many papers, in particular \cite{pontesch,Ip} in connection with the
applications to various quantum groups. The analogue of the $1/\cosh^2x$
potential in discrete QM with real shifts \cite{SZ} was reported in \cite{vdK}.

\section*{Acknowledgements}
We thank K.\,Aomoto for useful comments. R.\,S. thanks Pauchy Hwang and
Dept.$\!$\, Phys.$\!$\, National Taiwan University for hospitality.
S.\,O. is supported in part by Grant-in-Aid for Scientific Research
from the Ministry of Education, Culture, Sports, Science and Technology
(MEXT), No.25400395.

\bigskip\bigskip
\noindent{\LARGE\bf Appendix}
\appendix
\section{Some supporting evidences for the Conjecture}
\label{sec:app}

In this Appendix we provide a few supporting evidences for the
conjectured connection formula \eqref{phicon} and the replacement rule
between the $q$-shifted factorials and quantum dilogarithm functions
\eqref{trrule}.

First we note that
\begin{equation}
  \Phi^{(+)}_{\frac{\gamma}{2}}(z)\Phi^{(+)}_{\frac{\gamma}{2}}(-z-i\gamma)
  =\frac{\exp\bigl(\frac{i}{2\gamma}(z+i\frac{\gamma}{2}+i\pi)^2
  +\frac{i}{24\gamma}(\gamma^2+4\pi^2)\bigr)}
  {1-e^{-\frac{2\pi}{\gamma}z}}.
\end{equation}
By using this, we can show that \eqref{phicon} is consistent with twice
applications. Namely, applying \eqref{phicon} to the r.h.s.$\!$\, of
\eqref{phicon}, we have
\begin{align*}
  {}_2\phi_1\bigl(\genfrac{}{}{0pt}{1}
  {e^{\lambda},\,e^{\mu}}{e^{\nu}}|q\,;e^z\bigr)
  &=\ \ \,(\cdots)\Bigl((\cdots)\cdot{}_2\phi_1\bigl(\genfrac{}{}{0pt}{1}
  {e^{\lambda},\,e^{\mu}}{e^{\nu}}|q\,;e^z\bigr)
  +(\cdots)\cdot{}_2\phi_1\bigl(\genfrac{}{}{0pt}{1}
  {qe^{\lambda-\nu},\,qe^{\mu-\nu}}{q^2e^{-\nu}}|q\,;e^z\bigr)\Bigr)\n
  &\quad+(\cdots)\Bigl((\cdots)\cdot{}_2\phi_1\bigl(\genfrac{}{}{0pt}{1}
  {e^{\lambda},\,e^{\mu}}{e^{\nu}}|q\,;e^z\bigr)
  +(\cdots)\cdot{}_2\phi_1\bigl(\genfrac{}{}{0pt}{1}
  {qe^{\lambda-\nu},\,qe^{\mu-\nu}}{q^2e^{-\nu}}|q\,;e^z\bigr)\Bigr)\n
  &={}_2\phi_1\bigl(\genfrac{}{}{0pt}{1}
  {e^{\lambda},\,e^{\mu}}{e^{\nu}}|q\,;e^z\bigr).
\end{align*}
The corresponding result for the original connection formula \eqref{watform}
can be proven by using Riemann identity for the theta functions.

\noindent
\underline{$a=q^{-n}$ ($n\in\mathbb{Z}_{\ge0}$) case}:
When $a=q^{-n}$ ($n\in\mathbb{Z}_{\ge0}$) in the original connection
formula \eqref{watform}, the second term on the right hand side vanishes,
$(q^{-n};q)_\infty=0$ and it reduces to ((0.6.19) of \cite{koeswart} review):
\begin{equation}
  {}_2\phi_1\Bigl(\genfrac{}{}{0pt}{}{q^{-n},\,b}{c}\!\!\Bigm|\!q\,;z\Bigr)
  =\frac{(b;q)_n}{(c;q)_n}\,q^{-\frac12n(n+1)}(-z)^n\cdot
  {}_2\phi_1\Bigl(\genfrac{}{}{0pt}{}{q^{-n},\,q^{1-n}/c}
  {q^{1-n}/b}\!\!\Bigm|\!q\,;\frac{cq^{1+n}}{bz}\Bigr).
  \label{nwatform}
\end{equation}
Since both sides are finite sums, this formula is true for all complex $q$,
except for 0 and the root of unities ($q^m=1$, $1\leq m\leq n$).
In this case, the conjectured connection formula
\eqref{phicon} gives rise to the same formula as above. For this, we note
that $\Phi_{\frac{\gamma}{2}}^{(+)}({\lambda})$ in the denominator of the
second term on r.h.s.$\!$\, has a pole at $\lambda=e^{in\gamma}$ and the quantum
dilogarithms of the first term on r.h.s.$\!$\, give the correct factors:
\begin{align}
  &\frac{\Phi_{\frac{\gamma}{2}}^{(+)}({z})
  \Phi_{\frac{\gamma}{2}}^{(+)}({-i\gamma-z})}
  {\Phi_{\frac{\gamma}{2}}^{(+)}({z+in\gamma})
  \Phi_{\frac{\gamma}{2}}^{(+)}({-i\gamma-in\gamma-z})}
  =q^{-\frac12n(n+1)}(-e^z)^n,\\
  &\frac{\Phi_{\frac{\gamma}{2}}^{(+)}({\nu})
  \Phi_{\frac{\gamma}{2}}^{(+)}(\mu-in\gamma)}
  {\Phi_{\frac{\gamma}{2}}^{(+)}({\nu-in\gamma})
  \Phi_{\frac{\gamma}{2}}^{(+)}({\mu})}
  =\prod_{j=0}^{n-1}\frac{1-e^{\mu-ij\gamma}}{1-e^{\nu-ij\gamma}}
  =\frac{(e^{\mu};q)_n}{(e^{\nu};q)_n}.
\end{align}

\noindent
\underline{$q$-analogue of Euler's transformation formula}:
\begin{equation*}
  {}_2F_1\Bigl(\genfrac{}{}{0pt}{}{a,\,b}{c}\!\!\Bigm|\!z\Bigr)
  =(1-z)^{c-a-b}\cdot
  {}_2F_1\Bigl(\genfrac{}{}{0pt}{}{c-a,\,c-b}{c}\!\!\Bigm|\!z\Bigr),
\end{equation*}
reads for $0<q<1$ ((10.10.2) of \cite{askey})
\begin{equation}
  {}_2\phi_1\Bigl(\genfrac{}{}{0pt}{}{a,\,b}{c}\!\!\Bigm|\!q\,;z\Bigr)
  =\frac{(abz/c;q)_\infty}{(z;q)_\infty}\cdot
  {}_2\phi_1\Bigl(\genfrac{}{}{0pt}{}{c/a,\,c/b}
  {c}\!\!\Bigm|\!q\,;\frac{abz}{c}\Bigr).
  \label{qeuler}
\end{equation}
The replacement rule \eqref{trrule} says that the $|q|=1$ counterpart is
\begin{equation}
  {}_2\phi_1\Bigl(\genfrac{}{}{0pt}{}{e^{\lambda},\,e^{\mu}}{e^{\nu}}
  \!\!\Bigm|\!q\,;e^z\Bigr)
  =\frac{\Phi_{\frac{\gamma}{2}}^{(+)}({z})}
  {\Phi_{\frac{\gamma}{2}}^{(+)}({z+\lambda+\mu-\nu})}\cdot
  {}_2\phi_1\Bigl(\genfrac{}{}{0pt}{}{e^{\nu-\lambda},\,e^{\nu-\mu}}
  {e^{\nu}}\!\!\Bigm|\!q\,;e^{z+\lambda+\mu-\nu}\Bigr).
  \label{qeuler2}
\end{equation}
One can show that the r.h.s.$\!$\, satisfies the same difference equation
\eqref{q-difeq} as the l.h.s. At $z\to-\infty$, the basic hypergeometric
functions and the factors go to unity and the equality holds.
This is a supporting evidence for the replacement rule \eqref{trrule}.

\goodbreak

\end{document}